\documentclass[useAMS,usenatbib,a4paper]{mn2e}

\usepackage{graphicx,epsfig,natbib}
\usepackage{amssymb,txfonts}
\usepackage{epstopdf}

\newcommand{\de}{\mbox{d}}

\title[Stellar disruption by a SMBH]{Stellar disruption by a supermassive
  black hole: is the light curve really proportional to $t^{-5/3}$?}
\author[Lodato, King \& Pringle]{G. Lodato$^1$,  A. R. King$^1$ and J. E. Pringle$^{1,2}$\\
  $^1$ Department of Physics and Astronomy, University of Leicester, Leicester, LE1 7RH\\
  $^2$Institute of Astronomy, University of Cambridge, Madingley Road,
  Cambridge CB3 0HA } \date{} 

\begin{document}
\maketitle

\begin{abstract}
  In this paper we revisit the arguments for the basis of the time evolution
  of the flares expected to arise when a star is disrupted by a supermassive
  black hole. We present a simple analytic model relating the lightcurve to
  the internal density structure of the star. We thus show that the standard
  lightcurve proportional to $t^{-5/3}$ only holds at late times. Close to the
  peak luminosity the lightcurve is shallower, deviating more strongly from
  $t^{-5/3}$ for more centrally concentrated (e.g. solar--type) stars. We test
  our model numerically by simulating the tidal disruption of several stellar
  models, described by simple polytropic spheres with index $\gamma$. The
  simulations agree with the analytical model given two considerations. First,
  the stars are somewhat inflated on reaching pericentre because of the
  effective reduction of gravity in the tidal field of the black hole. This is
  well described by a homologous expansion by a factor which becomes smaller
  as the polytropic index becomes larger. Second, for large polytropic indices
  wings appear in the tails of the energy distribution, indicating that some
  material is pushed further away from parabolic orbits by shocks in the tidal
  tails. In all our simulations, the $t^{-5/3}$ lightcurve is achieved only at
  late stages. In particular we predict that for solar type stars, this
  happens only after the luminosity has dropped by at least two magnitudes
  from the peak. We discuss our results in the light of recent observations of
  flares in otherwise quiescent galaxies and note the dependence of these
  results on further parameters, such as the star/hole mass ratio and the
  stellar orbit
\end{abstract}

\section{Introduction}

X-ray flares from quiescent (non-AGN) galaxies are often interpreted as
arising from the tidal disruption of stars as they get close to a dormant
supermassive black hole (SMBH) in the centre of the galaxy \citep{komossa99}.
Similar processes also occur on much smaller scales, such as in compact binary
systems, where the black hole is only of stellar mass
\citep{rosswog08,rosswog08b}.

The pioneering work by \citet{lacy82,rees88,phinney89b} and, from a numerical
point of view, by \citet{evans89} have set the theoretical standard for the
interpretation of such events. In particular, a distinctive feature of this
theory is an apparent prediction for the time dependence of the light curve of
such events, in the form $L(t)\propto t^{-5/3}$ (note that the original paper
by \citealt{rees88} quotes a $t^{-5/2}$ dependence, later corrected to
$t^{-5/3}$ by \citealt{phinney89b}). Since then, a $t^{-5/3}$ light curve is
generally fitted to the observed luminosities of events interpreted as stellar
disruptions. In this paper, we revisit the theoretical arguments behind such
scaling and we show (as also originally argued by \citealt{rees88}) that the
light curve does {\it not need} to have this scaling and, in particular, that
it critically depends on the internal structure of the star being disrupted.
We provide a simple model to calculate the light curve starting from the
density profile of the star and we show that more centrally concentrated stars
tend to produce shallower light curves. We further supplement our model by a
numerical calculation of the process, using Smoothed Particle Hydrodynamics
(SPH).

We start by briefly summarizing the main argument of \citet{rees88}. Let us
consider a star, originally in hydrostatic equilibrium at a large distance
from the black hole. Since pressure forces and the internal self-gravity of
the star are in equilibrium, the only unbalanced force is the gravitational
pull of the black hole. The various fluid elements of the star therefore move
in essentially Keplerian orbits around the black hole, each one with its own
eccentricity, that is initially very close to the eccentricity of the centre
of mass of the star (in the following, for simplicity, we make the simple
assumption that the centre of mass of the star is in a parabolic orbit around
the black hole). Therefore, the distribution of specific mechanical energy
within the star is very narrow around the energy of the centre of mass.  As
the star moves closer to the black hole, the various Keplerian orbits tend to
be squeezed, perturbing the hydrostatic balance. Pressure forces then
redistribute energy inside the star, therefore widening the specific energy
distribution. After the encounter, the star is thus characterized by a much
wider distribution of internal energies, with part of the fluid having a
negative energy (and therefore being bound to the black hole) and part having
a positive energy (and therefore remaining unbound). In the picture of
\citet{rees88} it is this energy distribution (and only this) that determines
the light curve of the event. Indeed, after the encounter the fluid elements
again move in Keplerian orbits (but with their new energy). The bound elements
then come back close to pericentre after a Keplerian period $T$, linked to
their (negative) energy $E$ by:
\begin{equation}
E=-\frac{1}{2}\left(\frac{2\pi GM_{\rm h}}{T}\right)^{2/3},
\label{eq:ET}
\end{equation}
where $M_{\rm h}$ is the black hole mass. The mass distribution with specific
energy $\de M/\de E$ then translates, through Eq. (\ref{eq:ET}), into a mass
distribution of return times $\de M/\de T$. The next fundamental assumption is
that once the bound material has come back to the pericentre it loses its
energy and angular momentum on a timescale much shorter than $T$, thus
suddenly accreting onto the SMBH and giving rise to the flare. The mass
distribution of return times is therefore effectively the mass accretion rate
of the black hole during the event, from which the luminosity can be easily
computed. We thus have:
\begin{equation}
  \frac{\de M}{\de T}=\frac{\de M}{\de E}\frac{\de E}{\de T}=\frac{(2\pi GM_{\rm h})^{2/3}}{3}\frac{\de M}{\de E} T^{-5/3}.
\label{eq:MT}
\end{equation} 
In order to obtain the `standard' $t^{-5/3}$ light curve, we then have to make
the second fundamental assumption that the energy distribution is uniform.
Note that \citet{rees88} did not show that this should be the case, and only
assumed it for simplicity. Later, the numerical simulations by \citet{evans89}
apparently showed a uniform energy distribution, hence suggesting that the
light curve is generally proportional to $t^{-5/3}$. In the following, we
first show analytically that the energy distribution need not be uniform, but
depends on the properties of the star, and in particular on its internal
structure. We then show numerically that in fact it is not uniform and does
depend on the properties of the stars, in a way that approximately reproduces
the analytical results.

Starting from the pioneering work of \citet{carter82,carter83}, numerical
simulations of this process have been performed in a variety of studies
\citep{bicknell83,evans89,laguna93a,laguna93b,ayal2000,bogdanovic04}. Most of
these studies used Smoothed Particle Hydrodynamics (SPH,
\citealt{monaghan92}), mostly because of its ability of following the system
over a wide range of physical scales, where much of the simulated region is
essentially 'empty'. These various attempts have considered the effects of
varying the orbital parameters of the encounter \citep{bicknell83,evans89}, of
the inclusion of relativistic terms in the equation of motion
\citep{laguna93a,laguna93b,ayal2000} and have described the expected
observational outcome \citep{bogdanovic04}. However, surprisingly, no attempt
has been made at exploring the effect of varying the internal structure of the
star. Indeed, all such analyses have considered a simple polytropic model for
the star, with $\gamma$ invariably set to $5/3$ (although
\citealt{rosswog08,rosswog08b} has consider the encounter of a white dwarf
with a stellar mass black hole, which is in the much smaller mass ratio
regime).

The paper is organized as follows. In section 2 we describe our analytical
model to derive the energy distribution of the disrupted debris. In section 3
we describe our numerical code and the set up of our simulations. In section 4
we describe the results of the simulations. In section 5 we discuss our
results and draw our conclusions.

\section{The process of tidal disruption of a star by a SMBH}

\begin{figure}
  \centerline{\epsfig{figure=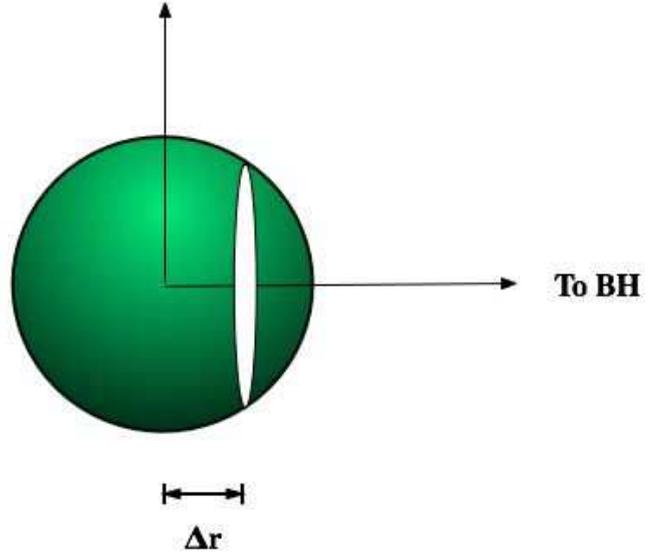,width=0.5\textwidth}}
  \caption{Schematic view of the geometry of the system. The radius of the
    star is $R_{\star}$The SMBH is on the right, at a distance $R_{\rm p}\gg
    R_{\star}$.}
\label{fig:scheme}
\end{figure}

\begin{figure*}
\centerline{\epsfig{figure=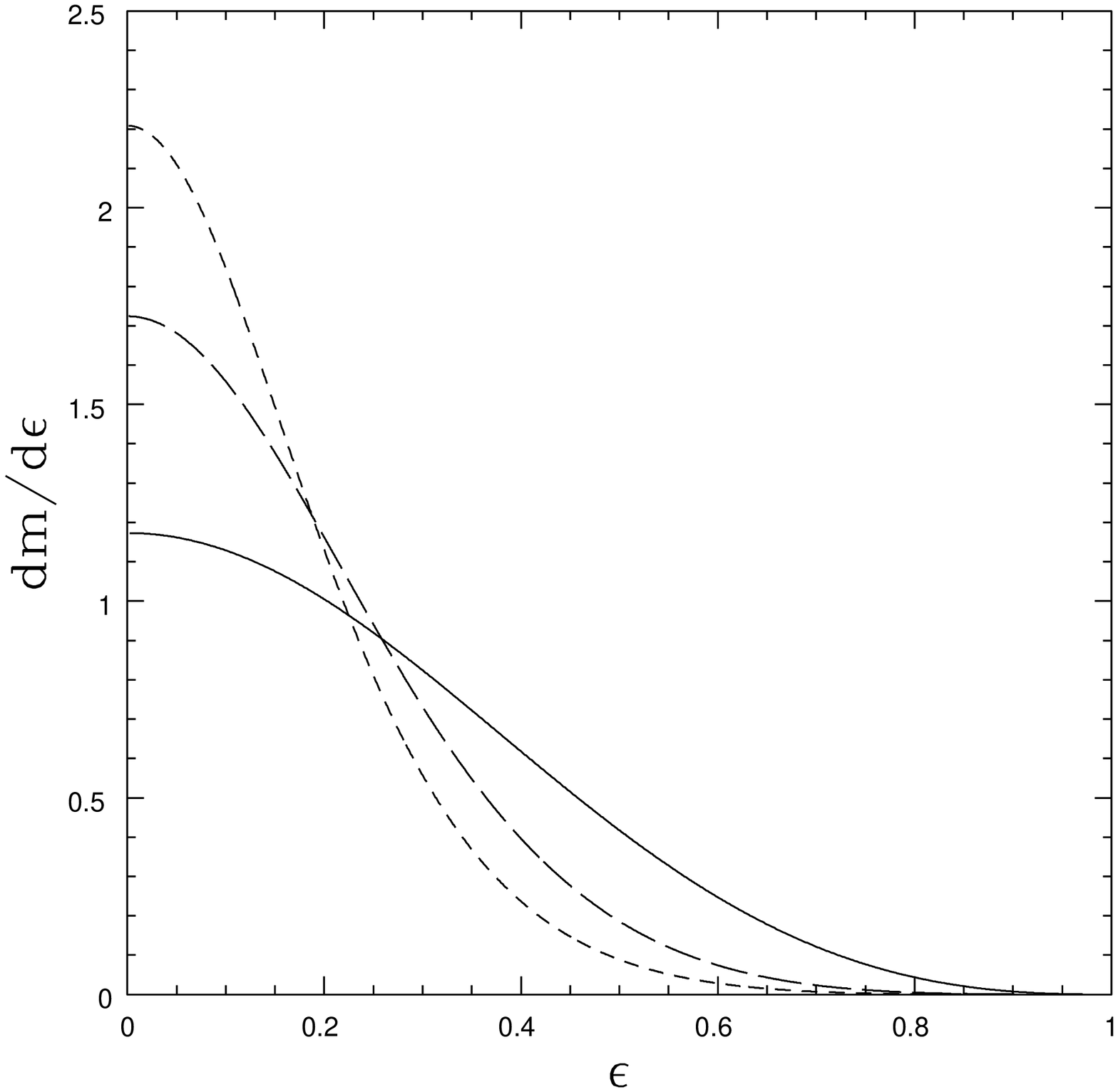,width=0.3\textwidth}
\epsfig{figure=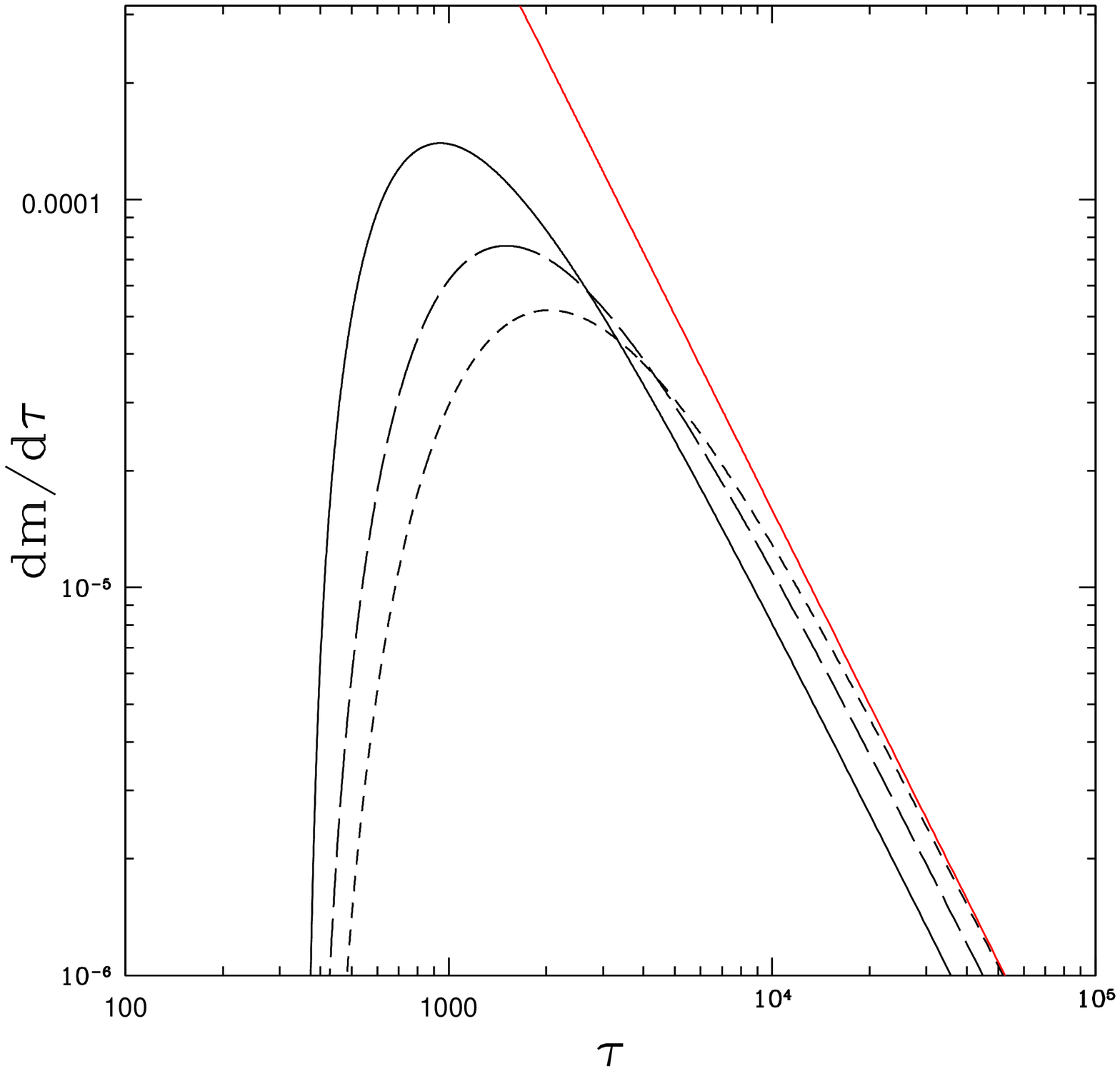,width=0.3\textwidth}
\epsfig{figure=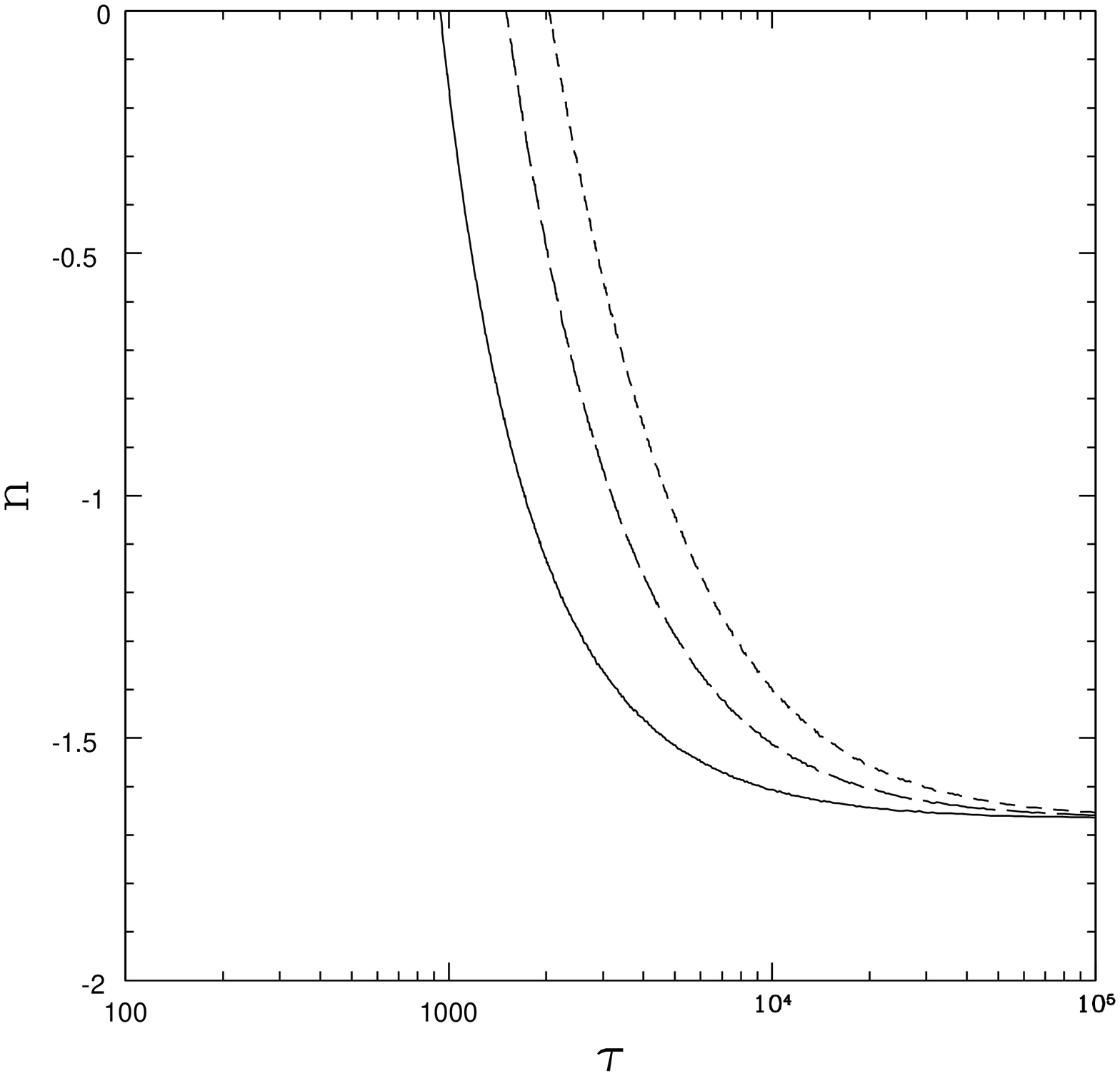,width=0.3\textwidth}}
\caption{Left: Distribution of internal energy for polytropic stars with
  different indices. Solid line: $\gamma=5/3$, long-dashed line: $\gamma =1.4$,
  short-dashed line: $\gamma=4/3$. Center: corresponding evolution of the
  accretion rate for the same three cases. The red line indicates for
  comparison a simple $t^{-5/3}$ power-law. Right: time evolution of the
  power-law index $n=\de\ln\dot{m}/\de\ln t$. As can be seen the value
  $n=-5/3$ is only approached at late times.}
\label{fig:analytic}
\end{figure*}

A simple and instructive way to consider the process is by treating the
interaction of the star with the black hole under the impulse approximation,
that is assuming that the interaction occurs in a very small time span as the
star gets close to pericentre. This approximation is probably appropriate for
highly hyperbolic encounters, but is only approximate for the parabolic case
considered here. We therefore expect the details to differ somewhat from what
is derived here below, even if the qualitative behaviour is correct.  In this
approximation, the motion of the star is simply a straight line until it
reaches pericentre, at which point it is subject to a short impulse that
deflects the various fluid elements, each of them individually conserving
their specific energy. Until it reaches pericentre, therefore, the structure
of the star is essentially unchanged and it keeps its original radial density
profile and its initial radius. The key thing to realize here is that the
spread of specific energy the star has reached just before the impulse is
simply given by the different depths at which the various fluid elements are
within the black hole potential well, because they all share the same velocity
and therefore the same kinetic energy. This was clearly realized by
\citet{lacy82} and \citet{evans89} who pointed out that ``The spread in
specific energy of the gas... is given by the change in the black hole
potential across a stellar radius''. The impulse occurs istantaneously and
therefore does not modify the kinetic energy of the fluid emlements but
imparts some degree of rotation in the star, with $\Omega\approx (2GM_{\rm
  h}/R_{\rm p}^3)^{1/2}$, where $R_{\rm p}$ is the pericenter distance (cf.
\citealt{evans89}). We consider here the case in which the stellar radius
$R_{\star}$ is much smaller than $R_{\rm p}$, which is appropriate for the
case of stellar disruption by a supermassive black holes (although not for the
case of compact binaries). This allows us to easily estimate the expected
energy spread, as
\begin{equation}
\Delta E = \left(\frac{\de E_{\rm p}}{\de r}\right)_{R_{\rm p}}\Delta r_{\rm max} = \frac{GM_{\rm h}}{R_{\rm p}^2}R_{\star},
\label{eq:deltaE}
\end{equation}
where $E_{\rm p}=GM_{\rm h}/r$ is the potential energy due to the black hole
and $\Delta r_{\rm max}= R_{\star}$ is the maximum deviation from the
pericentre distance (cf. \citealt{lacy82}). We thus expect the energy
distribution to extend roughly between $-\Delta E$ and $\Delta E$. This simple
result has been obtained in all the early analyses of the problem. However, in
the same approximation as before, we are also able to derive the whole energy
distribution starting from the density, by calculating what is the fraction of
stellar mass at a given $\Delta r$ from the centre. Fig. \ref{fig:scheme}
illustrates the geometry. It can be easily shown that
\begin{equation}
\frac{\de M}{\de\Delta r} = 2\pi \int_{\Delta r}^{R_{\star}}\rho(r)r\de r,
\label{eq:MR}
\end{equation}
where $\rho(r)$ is the spherically symmetric mass density of the star. The
relation between the distribution of $\Delta r$ and the distribution of energy
$E$ is simply given by:
\begin{equation}
\frac{\de M}{\de E} = \frac{\de M}{\de\Delta r}\frac{R_{\star}}{\Delta E},
\label{eq:ME}
\end{equation}
where $\Delta E$ is given by Eq. (\ref{eq:deltaE}).

It is useful to introduce dimensionless quantities. We then define $\epsilon =
-E/\Delta E$ (where we have also included a minus sign because we are
interested in material with negative specific energy) as our dimensionless
energy, $x=\Delta r/R_{\star}$ as our radial coordinate within the star,
$x_{\rm p}=R_{\rm p}/R_{\star}\gg 1$ as our dimensionless pericentre distance
and $m=M/M_{\star}$ as our dimensionless mass. We also introduce a fiducial
time unit $T_0 = 2\pi(R_{\rm p}^3/GM_{\rm h})^{1/2}$ and a dimensionless time
$\tau = T/T_0$, as well as a fiducial density $\rho_0=M_{\star}/R_{\star}^3$
and a dimensionless density $\hat{\rho}=\rho/\rho_0$. The outcome of the
disruption event depends on the 'penetration factor' $\beta=R_{\rm p}/R_{\rm
  t}$, that is the ratio of the pericenter distance to the tidal radius
$R_{\rm t}=q^{1/3}R_{\star}$, where $q=M_{\rm h}/M_{\star}$ is the mass ratio
between the black hole and the star. In order to tidally disrupt the star we
require $\beta\lesssim 1$. For example, for a mass ratio $q= 10^6$, we have
$\beta=1$ for $R_{\rm p}=100 R_{\star}$.  To give an idea of the numbers
involved, we note that for $R_{\rm p}=100R_{\odot}$, $M_{\rm
  h}=10^6M_{\odot}$, $M_{\star}=1M_{\odot}$ and $R_{\star}=R_{\odot}$, we have
$T_0 \approx 3.18~10^{-4}$ yrs $\approx 0.11$ days, while the unit for the
accretion rate is $M_{\star}/T_0\approx 3.1~10^3M_{\odot}$/yr.  In these
units, Eqs. (\ref{eq:ET}), (\ref{eq:MT}), (\ref{eq:MR}) and (\ref{eq:ME})
become simply:
\begin{equation}
\epsilon = \frac{x_{\rm p}}{2}\tau^{-2/3},
\label{eq:dim1}
\end{equation}
\begin{equation}
\frac{\de m}{\de\tau} = \frac{x_{\rm p}}{3}\frac{\de m}{\de\epsilon}\tau^{-5/3},
\label{eq:dim2}
\end{equation}
\begin{equation}
\frac{\de m}{\de x} = 2\pi\int_x^1\hat{\rho}(x')x'\de x',
\label{eq:dim3}
\end{equation}
\begin{equation}
\frac{\de m}{\de \epsilon}=\frac{\de m}{\de x} .
\label{eq:dim4}
\end{equation}

The above simple set of equations therefore allows us to calculate the
accretion rate onto the black hole as a function of the internal stellar
structure. In general, we expect the density to show a peak at small radii $x$
and therefore a peak at small specific energies $\epsilon$. Since material at
lower energies contributes to the accretion at later times, we can already
predict what relative changes do we expect with respect to the standard
$t^{-5/3}$ light curve. In particular, we expect that if the star is more
centrally condensed the flare should start with a relatively longer delay
(less matter at large energies - small return time) and should have a
shallower light curve (more matter at small energies - large return time).
However, unless the density is strongly diverging at small radii, we expect
$\de m/\de x= \de m/\de\epsilon$ to flatten at the lowest energies and
therefore the light curve to approach a $t^{-5/3}$ profile at late times.

\begin{figure*}
\centerline{\epsfig{figure=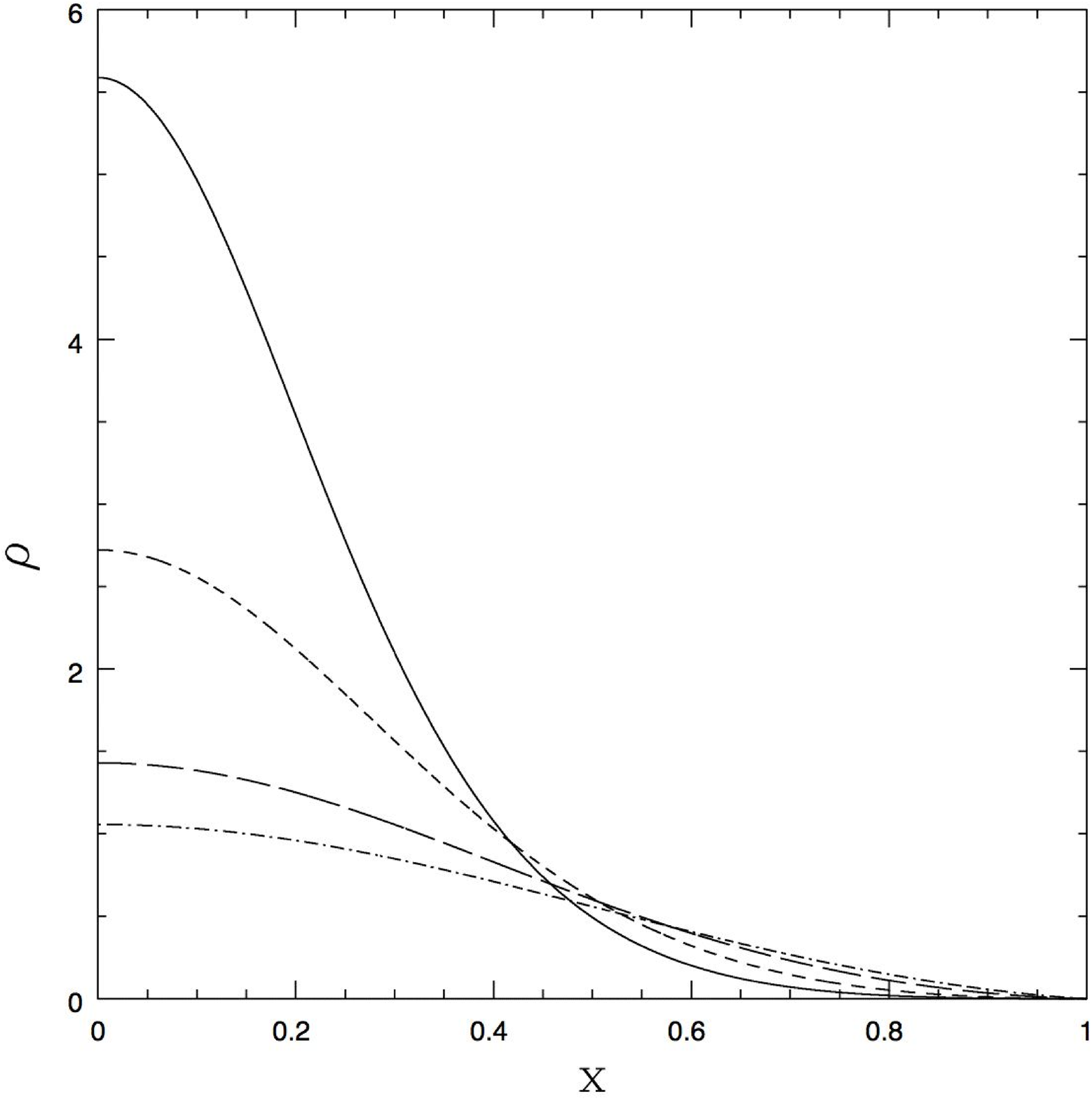,width=0.5\textwidth}
\epsfig{figure=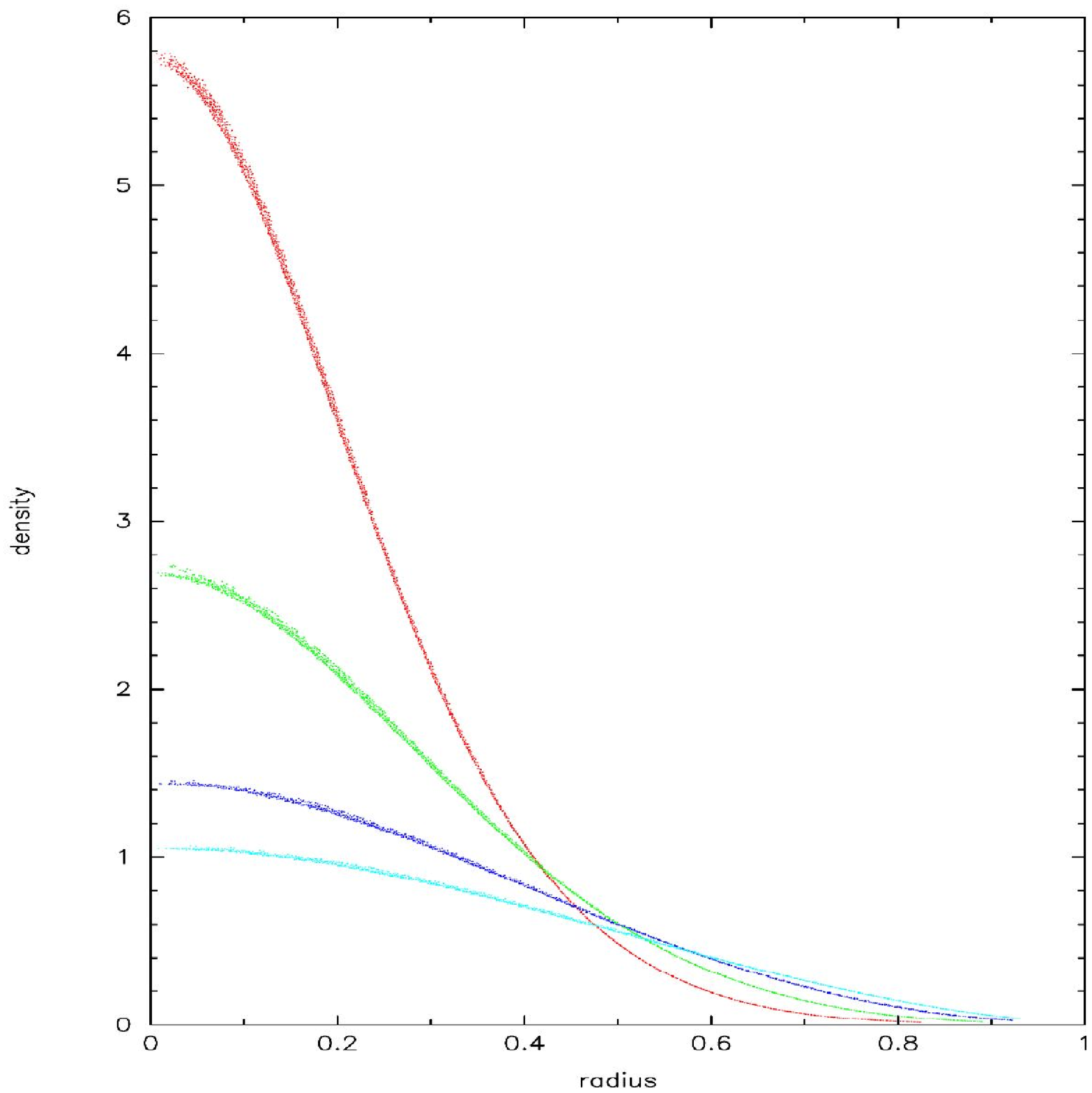,width=0.5\textwidth}
}
\caption{Radial density profiles for the four models considered here. In the
  left panel we show the density of four solutions of the Lane-Emden equation
  with (from the highest to the lowest central density) $\gamma=1.4$, 1.5, 5/3
  and 1.8. In the right panel we show the corresponding SPH density estimates
  for the initial conditions of our simulations.}
\label{fig:initial}
\end{figure*}

As an example, we can use the above analytical formulae to calculate the
specific energy distribution and the accretion rate as a function of time
predicted for some simple stellar models with known density profiles. We have
thus considered simple polytropic spheres with different indices $\gamma =
5/3$, 1.4 and $4/3$. We have first solved numerically the Lane-Emden equation
for the three cases and have then computed the various relevant quantities
using Eq. (\ref{eq:dim1})-(\ref{eq:dim4}) above, assuming $x_{\rm p}=100$. The
results are shown in Fig.  \ref{fig:analytic}.  The left panel shows the
prediction for the energy distribution, where the solid line indicates the
relatively non compact case $\gamma=5/3$, the short-dashed line indicates
$\gamma=1.4$ and the long-dashed case shows the most compact case
$\gamma=4/3$. As can be seen, the energy distributions do extend up to
$\epsilon \sim 1$, but are not flat except at very low energies, the effect
becoming more pronounced for the more compact cases. The middle panel shows
the predicted evolution of the mass accretion rate $\dot{m}=\de m/\de \tau$,
for the three values of $\gamma$ with the same line styles as the left panel.
The red line shows for comparison a simple power law with index -5/3. It can
be seen that indeed the light curves are slightly shallower that $t^{-5/3}$
and approach it only at late times. This is even more evident in the right
panel, where we plot the power law index $n=\de\ln\dot{m}/\de\ln\tau$ for the
three cases. If we want to put some numbers on the estimates above, note that
for our standard numerical values described above a time of 1 year corresponds
to roughly $\tau\approx 3000$. We then see that the power law index after 1
year of the flare is $n\approx -1.5$ for $\gamma=5/3$, which is reasonably
close to the expected -5/3. However, such stellar model is probably
unrealistic for a solar type star, whose structure is rather more similar to a
$\gamma=4/3$ polytrope, in which case, after 1 year of the flare the power law
index is still $n\approx -0.8$.

\section{Numerical simulations}

The model described in the previous section is only approximate in that it
treats the interaction between the star and the black hole as instantaneous.
In particular, the distribution of specific energy of the disrupted stellar
material, and consequently the resultant lightcurve, has been computed by
assuming that the stellar structure is essentially unchanged until it reaches
pericentre. Still, it highlights some important features of the stellar
disruption process: the expected energy distribution is in general not flat,
and it tends to become progressively more peaked towards lower energies as the
stellar structure model gets more centrally concentrated (that is, as the
polytropic index $\gamma$ becomes smaller). In order to gain a better
understanding of the process, we have therefore compared the analytical
expectations with the results of numerical hydrodynamical simulations of the
process.

\begin{figure*}
\centerline{\epsfig{figure=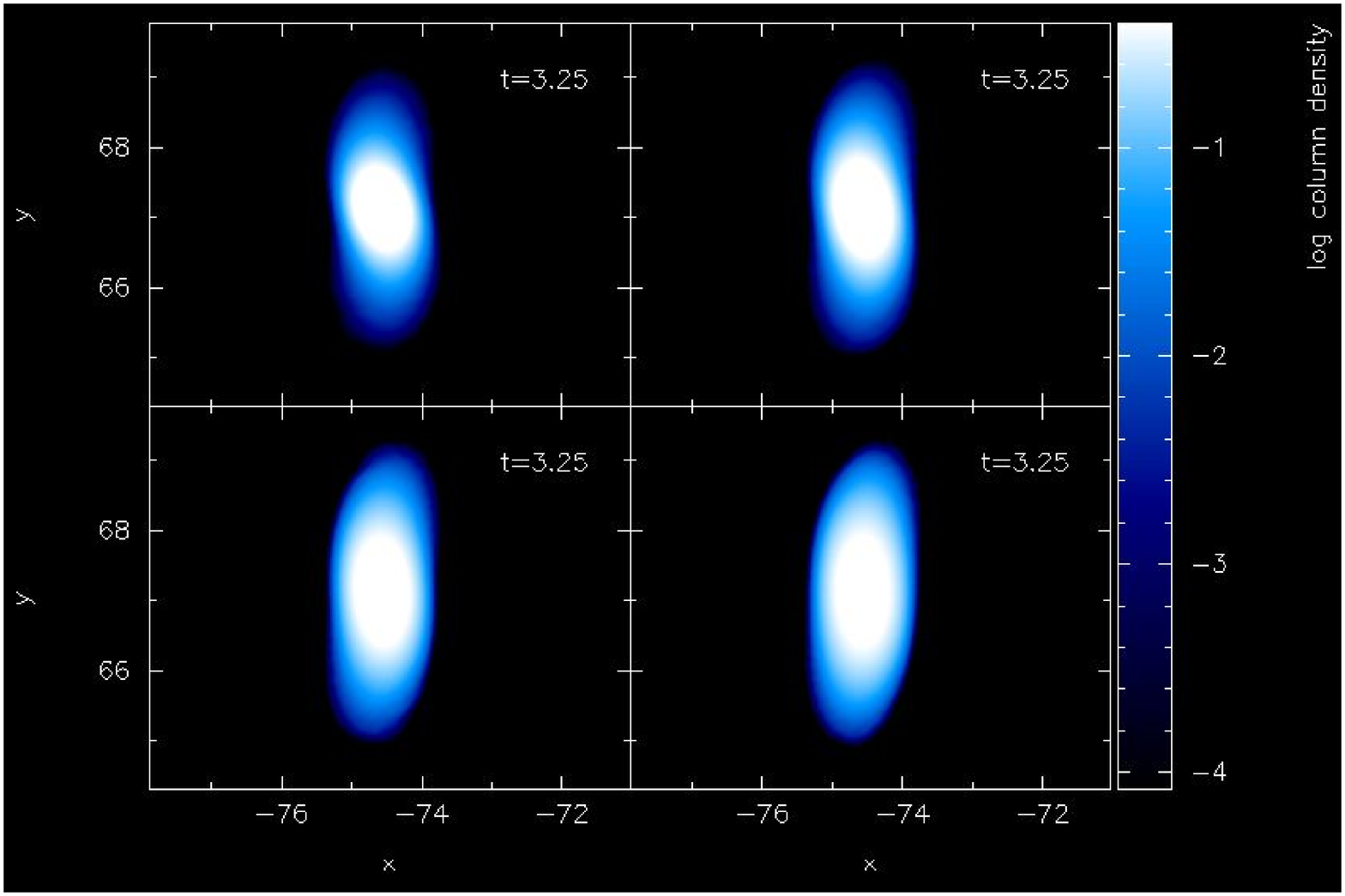,width=0.5\textwidth}
\epsfig{figure=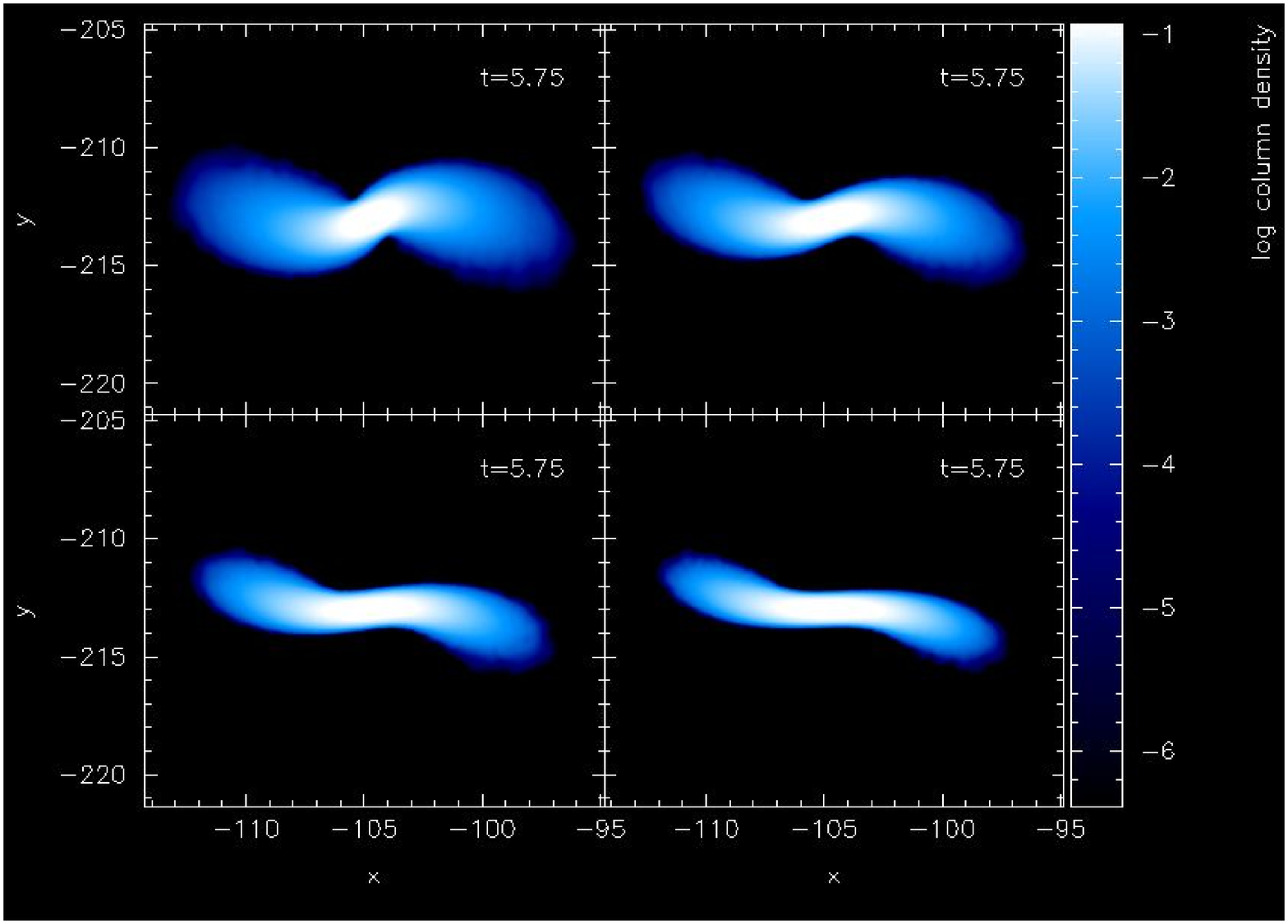,width=0.465\textwidth}
}
\caption{{\bf (a)} Projected density of the of the star at pericentre. 
The black hole is outside the image, at the origin of the
  coordinate system. The four panels refer to different values of $\gamma=$
  1.4 (upper left), 1.5 (upper right), 5/3 (lower left) and 1.8 (lower
  right). {\bf (b)} Same as panel {\bf (a)}, but after
  the encounter, when the star is located at roughly two times the pericentre
  distance.}
\label{fig:image}
\end{figure*}

\subsection{Numerical setup}

In the case where the encounter is parabolic, as mentioned above, the two
relevant dynamical parameters are the mass ratio between the star and the
black hole, $q=M_{\rm h}/M_{\star}$ and the penetration factor $\beta = R_{\rm
  p}/R_{\rm t}$. In this work we have considered the case where $\beta=1$ and
$q=10^6$, which imply that the pericentre distance is equal to 100 times the
radius of the star.

Following the several investigations summarized in the Introduction, we have
also used a non-relativistic SPH code to simulate the encounter. Our code uses
individual particle timesteps \citep{bate95}, it evolves the smoothing length
by keeping a fixed mass within a smoothing sphere (equivalent to roughly 60
particles) and includes the relevant terms needed to ensure energy
conservation when the smoothing length is variable (see \citealt{price05} for
a recent review). We also adopt a standard SPH artificial viscosity
\citep{monaghan92} with viscosity parameters $\alpha_{\rm sph}=1$ and
$\beta_{\rm sph}=2$.

In order to describe the basic dynamics of the encounter we do not require to
use an extremely large number of particles in order to reach a satisfactory
resolution. Indeed, \citet{evans89} have shown that their results were
numerically converged with a number of particles $N$ equal to a few $10^4$.
Even recent calculations have only used a relatively small number of
particles, of the order of $10^3$ \citep{ayal2000} up to $2~10^4$
\citep{bogdanovic04}. In this work we have run all our simulations at the two
resolution of $N=10^4$ and $N=10^5$ and have noticed no appreciable difference
in the results, thereby confirming the numerical convergence of the results.
In the following, we only show the higher resolution results.

We initialize our simulations by placing the SPH particles to form the
structure of a polytropic star of given index $\gamma$ (we have considered the
four cases $\gamma = 1.4$, 1.5, 5/3 and 1.8). This is done by initially
placing the particles using close sphere packing and then differentially
stretching their radial position to achieve the desired density profile. This
method minimizes the statistical noise associated with random placing of the
particles (we thank Walter Denhen for providing this setup routine). We then
relax the structure of the star by evolving it in isolation until its internal
properties settle down.

We have considered four different values of $\gamma=$ 1.4, 1.5, 5/3 and 1.8.
In this way we encompass the expected range for different kinds of stars, from
radiative to convective ones. Indeed, a solar type star has a density profile
close to a $\gamma=4/3$ polytrope (it is actually best described by
$\gamma\approx 1.3$). Unfortunately, a $\gamma=4/3$ polytrope is difficult to
simulate, as it has zero binding energy. The lowest value of $\gamma$ that we
use is then 1.4. Red giants and low mass stars can be described by a
$\gamma=5/3$ polytrope, while neutron stars have a structure which is probably
closer to a $\gamma=2.5$ polytrope.

We plot in Fig. \ref{fig:initial} the initial density
profile of our four models as predicted from the solution of the Lane-Emden
equation (left panel) and as realized after the initial conditions have been
allowed to relax (right). As can be seen, the four models differ in their
central concentration, such that the $\gamma=1.4$ model is the most
concentrated and the $\gamma=1.8$ is the least. It might be worth also to
recall that models with larger $\gamma$ are less compressible than models with
lower $\gamma$. 

Finally, we introduce the black hole as a point mass at the origin and we
displace the star so as to place its center of mass on the required parabolic
orbit (since the star is an extended object this actually means that the total
mechanical energy of the star is slightly negative, amounting to roughly
-0.005 in our units). The initial distance from the black hole is three times
the pericentre distance (in other simulations not described here, we have also
used a larger initial distance and found no significant difference). Our code
units are $R_{\star}$ for length and $M_{\star}$ for mass, which ensure that
our results are described in the same dimensionless variables as described in
Section 2.  The black hole is modelled as a sink onto which SPH particles can
be accreted if they come closer to the black hole that a distance 0.25 in code
units.  However, in practice, given that our pericentre is very large and that
we do not follow the evolution of the debris long after the interaction, no
particles are actually accreted during the course of our simulations.

\section{Results}

\subsection{The $\gamma=5/3$ case}

Before comparing the results obtained with various polytropic indices, we
start by describing the results that we have obtained in the $\gamma=5/3$,
which is directly comparable to the simulations discussed in previous papers.
In particular, this simulation is essentially a higher resolution version of
the one initially discussed in \citet{evans89}.

Two snapshots of the integrated density profile of the star are shown in the
lower left panels of Fig. \ref{fig:image}(a,b), at two different times, that
is when the star is at pericentre and when it is at roughly two times the
pericentre distance, after the encounter. The overall structure of the star
looks qualitatively similar to the one shown in \citet{evans89}. It is
interesting to notice that at pericentre the star is already quite distorted
with respect to its initial configuration and in particular it has expanded
somewhat (recall that its initial radius is 1 in code units). This occurs
because, in isolation, the star is in hydrostatic equilibrium between its
pressure and its self-gravity. As the star approaches the black hole the tidal
field effectively acts as to reduce the stellar gravity, making pressure
forces unbalanced and therefore `inflating' the star. This effect is expected
to be more significant for small than for large $\gamma$. This reflects the
fact that the radius of a polytrope with small $\gamma$ is more sensitive to
the effective gravity.  

\begin{figure}
\centerline{\epsfig{figure=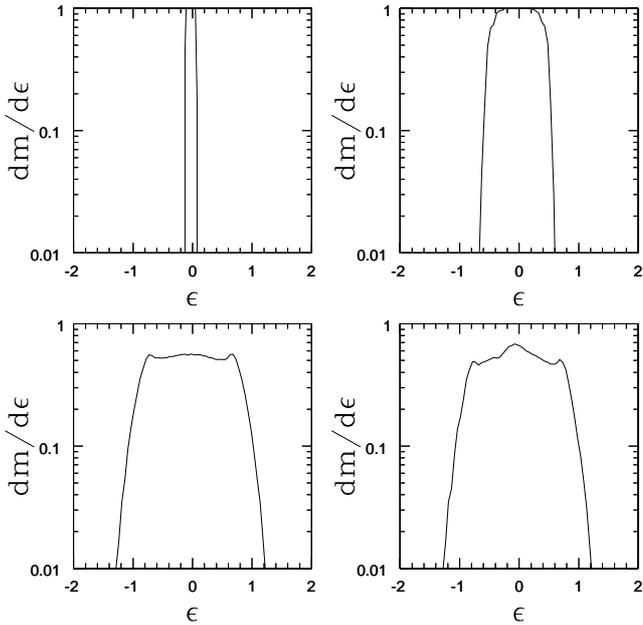,width=0.5\textwidth}
}
\caption{Distribution of specific energies for the case $\gamma=5/3$. The four
  show the distribution for the initial condition (upper left panel), when the
  star is at pericentre ((upper right), at four times the pericentre distance
  (lower left) and at ten times the pericentre distance (lower right).}
\label{fig:evans}
\end{figure}

A more quantitative comparison can be done by looking at the distribution of
specific energies of the disrupted star. This is shown in Fig. \ref{fig:evans}
at four different times during the simulations: at $t=0$ (upper left panel),
at pericentre (upper right), and after the encounter, when the star is roughly
at four times the pericentre distance (lower left) and ten times the
pericentre distance (lower right). For ease of comparison with
\citet{evans89}, only for this plot we have used a logarithmic scale for the
distribution. It can be seen that initially the distribution is very narrow
and centered at $\epsilon=0$, which just reflect the fact that the whole star
is initially on a parabolic orbit. As the star approaches the black hole, the
distribution becomes wider and indeed approaches the width predicted by the
simple analysis of Section 2 (which is equal to unity in the units adopted
here). The lower left panel, in particular, showing the distribution at four
times the pericentre, compares almost exactly with the distribution shown by
\citet{evans89} (their fig. 3), confirming that indeed our simulations
replicate accurately their results. However, one can see that the density
distribution keeps evolving until the star is at roughly 10 pericentre
distances, where it finally settles down in the configuration shown in the
lower right panel of Fig. \ref{fig:evans}. We thus see that the distribution
is characterized by a central peak at lower energies, followed by two `wings'
at larger energies. The presence of a central peak is expected based on the
analytical model described above. The wings, on the other hand, refer to the
stellar material at the surface of the star, which at pericentre is somewhat
distorted from its initially spherical shape (as can be seen in Fig.
\ref{fig:image}, lower left panel) and would obviously show some discrepancies
with respect to the simple `spherical' model of Section 2.

\begin{figure}
\centerline{\epsfig{figure=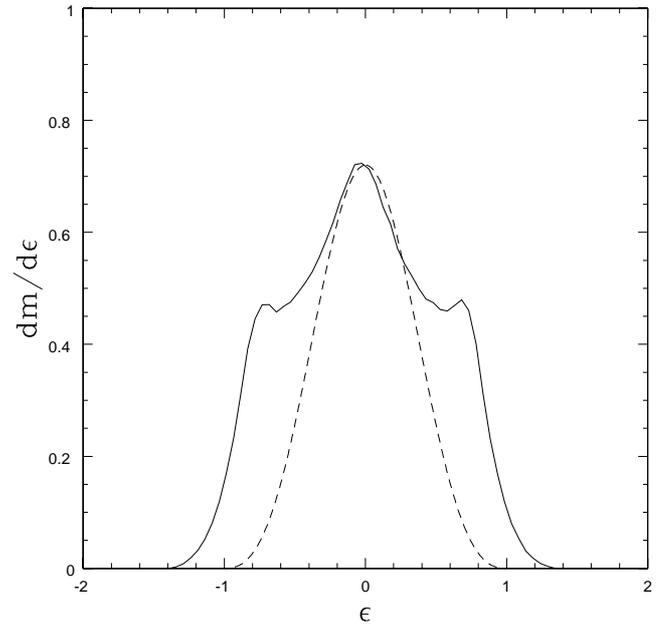,width=0.5\textwidth}
}
\caption{Distribution of specific energies for the case $\gamma=5/3$. Solid
  line: average distribution at the end of the simulation. Dashed line:
  predicted distribution based on the analytical model, re-normalized to match
  the peak.}
\label{fig:spec1}
\end{figure}

Fig. \ref{fig:spec1} (solid line) shows the distribution of specific energies
averaged over 10 time units, when the stars has reached $\sim 20$ pericentre
distances and the distribution has settled down. This is compared with the
prediction of the analytical model of Section 2 (cf. Fig.  \ref{fig:analytic},
left panel), which is shown with a dashed line. Since the profiles are all
normalized to 1, in order to compare the shape of the distribution at the
peak, we have scaled down the analytical profile by a factor $\approx 1.6$. We
thus see that the analytical profile does approximately match the shape of the
distribution at the peak, except for the presence of the wings, indicating the
presence of more material at extreme energies than predicted by the model.
Note that, obviously, the light curve produced in this case does not show the
standard $t^{-5/3}$ decline, especially at early times. More details on the
resultant light curves are given in the next Section, where we compare the
results obtained with different values of the polytropic index.

\subsection{Varying the polytropic index}

We now discuss the effects of varying the polytropic index $\gamma$ on the
structure of the disrupted star. A first comparison can be obtained by looking
at Fig. (\ref{fig:image}), where the various panels show the projected density
of the star at pericentre (left) and at two times the pericentre (right) for
the four cases considered (from top left to bottom right: $\gamma=1.4,$ 1.5,
5/3 and 1.8). Several interesting differences can be already seen from these
images. First of all, note that the overall expansion of the star is similar
in all cases. However, for larger values of $\gamma$ the density structure of
the star is much more uniform. This is particularly evident in the right
panel, which refers to well after pericentre passage. In the case where
$\gamma=1.4$ the high density core is compact, with the density in the `puffy'
tidal tails gently declining. In contrast, at the opposite extreme of
$\gamma=1.8$ the high density region is more extended and the edge of the
tidal tails is more clearly defined, revealing a sharper density cut-off at
the edge. It is also interesting to note the different degree of internal
rotation induced in the star by the tidal interaction, with the elongated core
being more aligned with the line joining the star and the black hole (at the
origin of the coordinate system) for smaller $\gamma$ than for larger ones.

\begin{figure*}
\centerline{\epsfig{figure=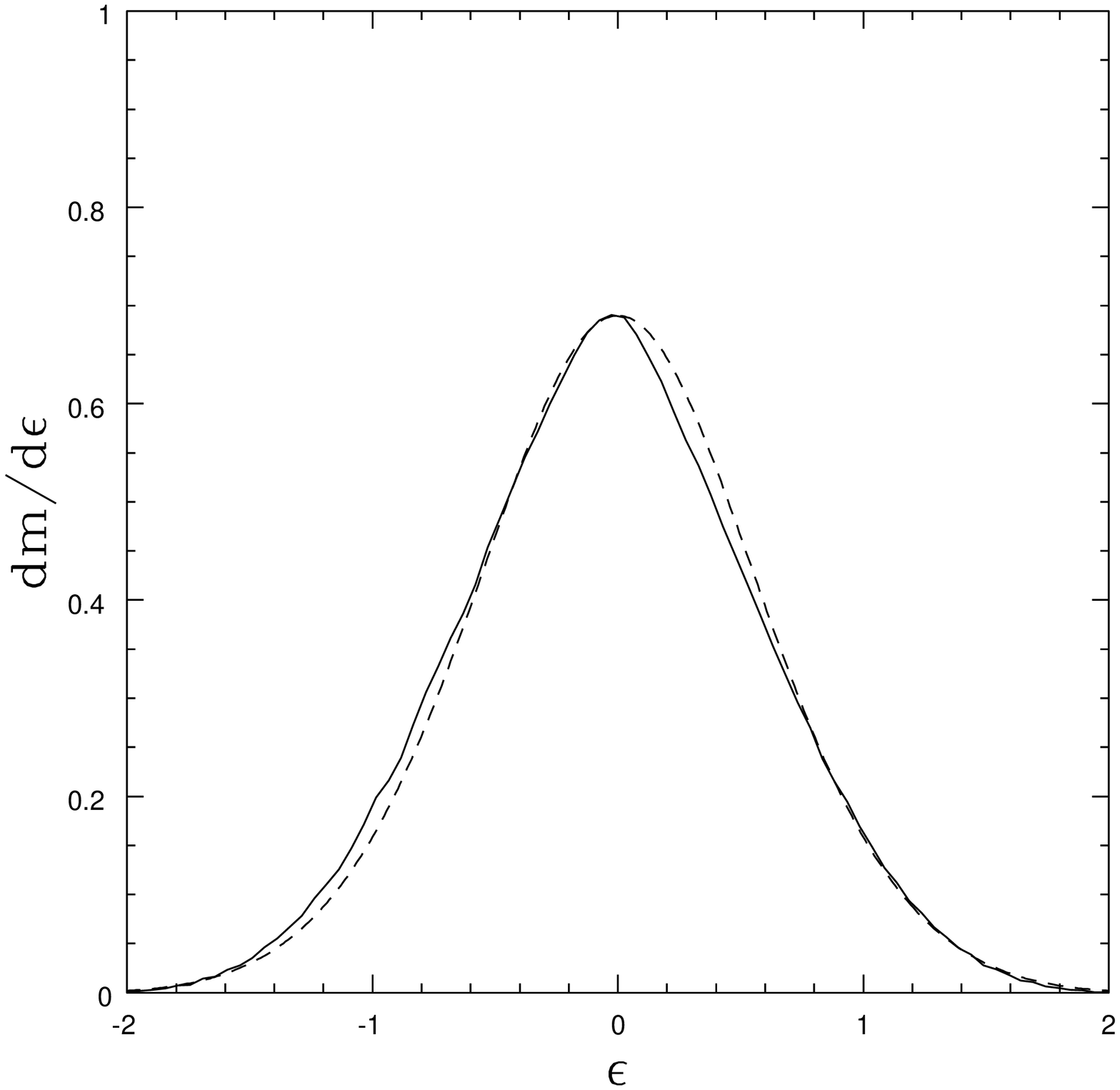,width=0.35\textwidth}
\epsfig{figure=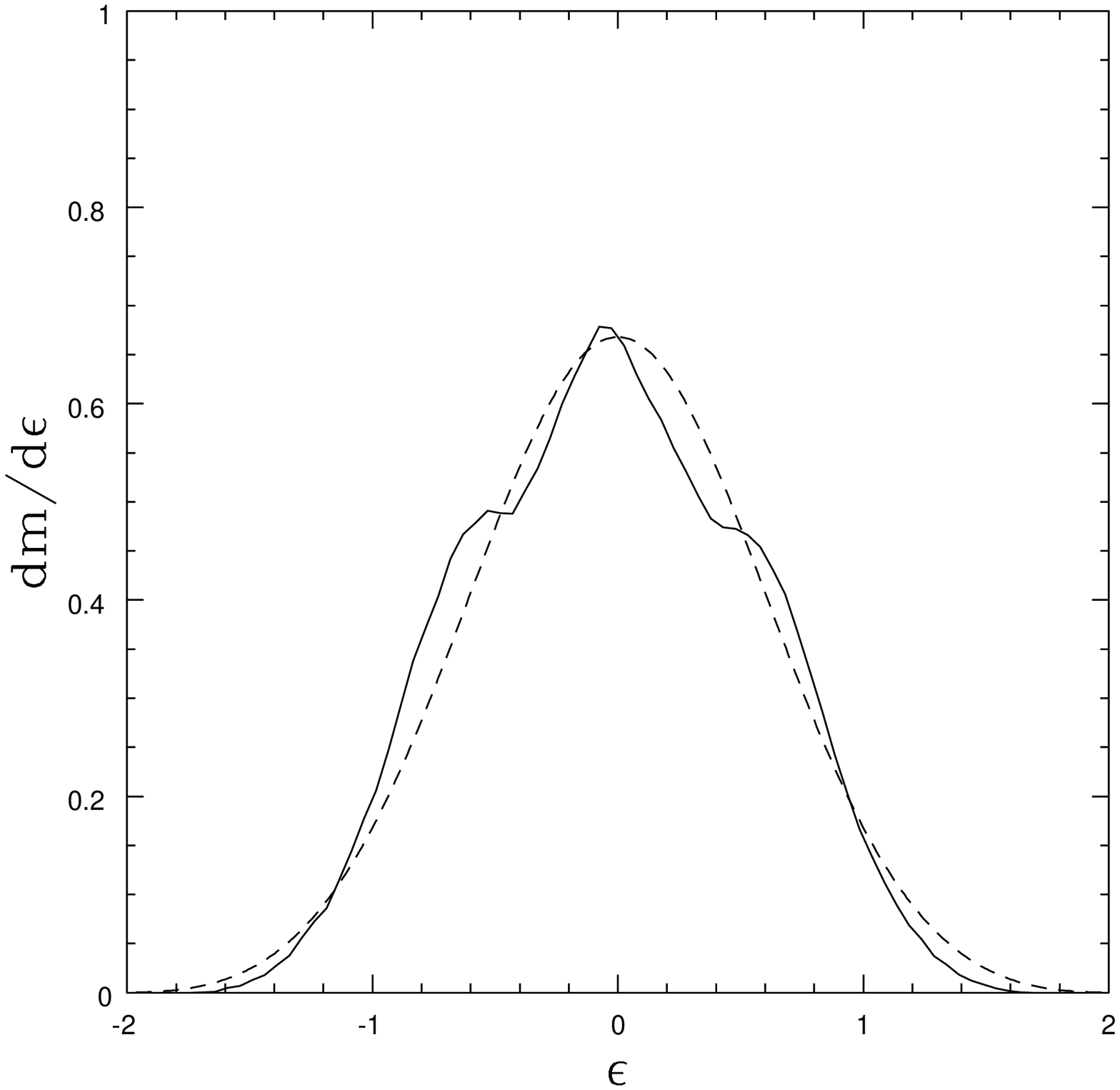,width=0.35\textwidth}}
\centerline{\epsfig{figure=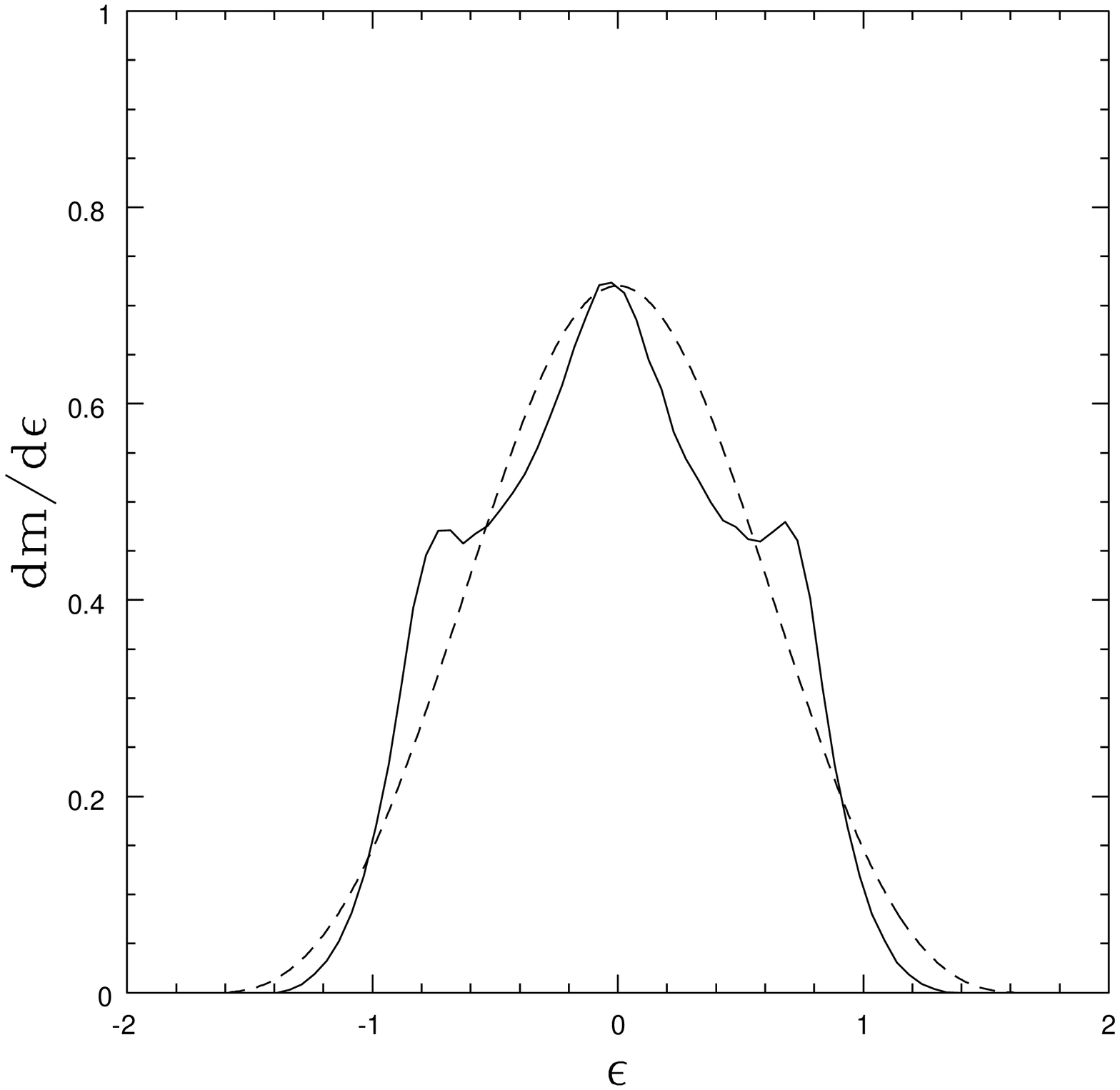,width=0.35\textwidth}
\epsfig{figure=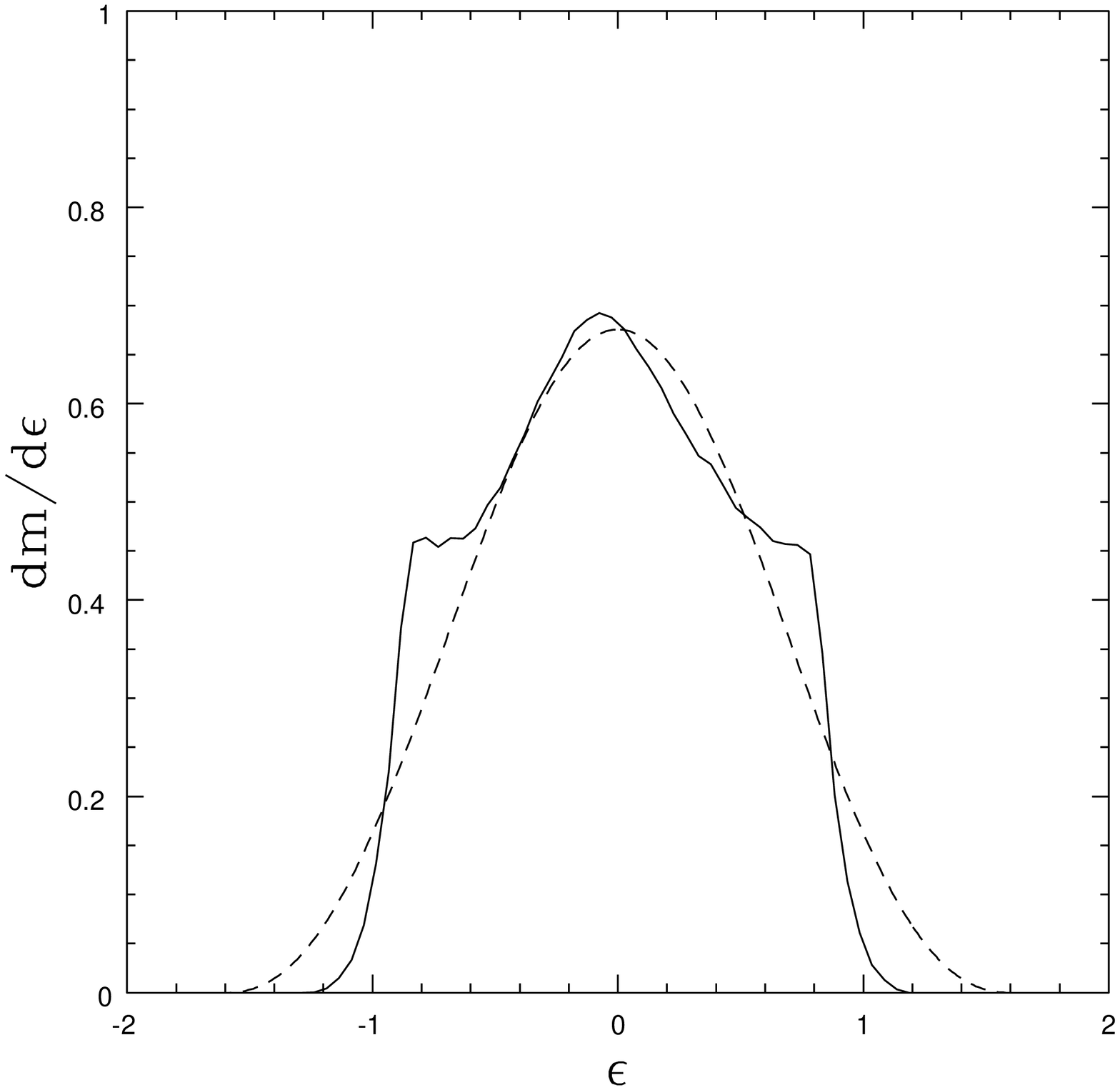,width=0.35\textwidth}}
\caption{Specific energy distribution for the four simulations. Upper left:
  $\gamma=1.4$, upper right: $\gamma=1.5$, lower left: $\gamma=5/3$, lower
  right: $\gamma=1.8$. The solid lines are the result of the simulations,
  while the dashed lines show the distribution expected from the simple
  analytical theory outlined in Section 2, where in each case the initial
  density profile of the star has been homologously expanded by a factor
  $\xi=2.5$, 2.1, 1.63 and 1.6 for the four cases $\gamma=1.4$, 1.5, 5/3 and
  1.8, respectively.}
\label{fig:specave}
\end{figure*}

\begin{figure}
\centerline{\epsfig{figure=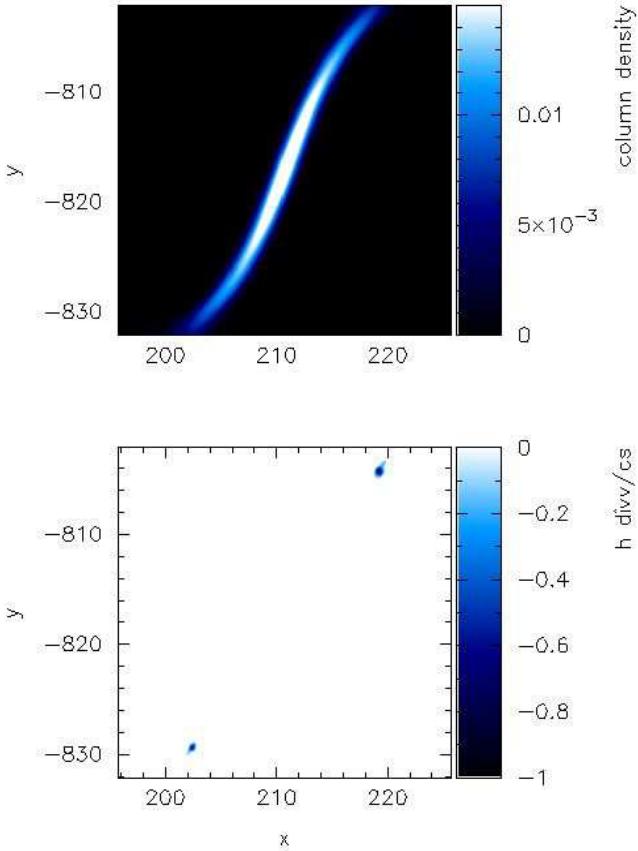,width=0.5\textwidth}}
\caption{Top: Projected density fot the $\gamma=5/3$ case at
  $t=16.75$. Bottom: vertical cross section of the quantity $q$, defined in
  Eq. (\ref{eq:q}). When $|q|>1$ the gas undergoes a shock. It can be seen
  that this occurs at the edge of the tidal tails.}
\label{fig:shockimage}
\end{figure}

\begin{figure}
\centerline{\epsfig{figure=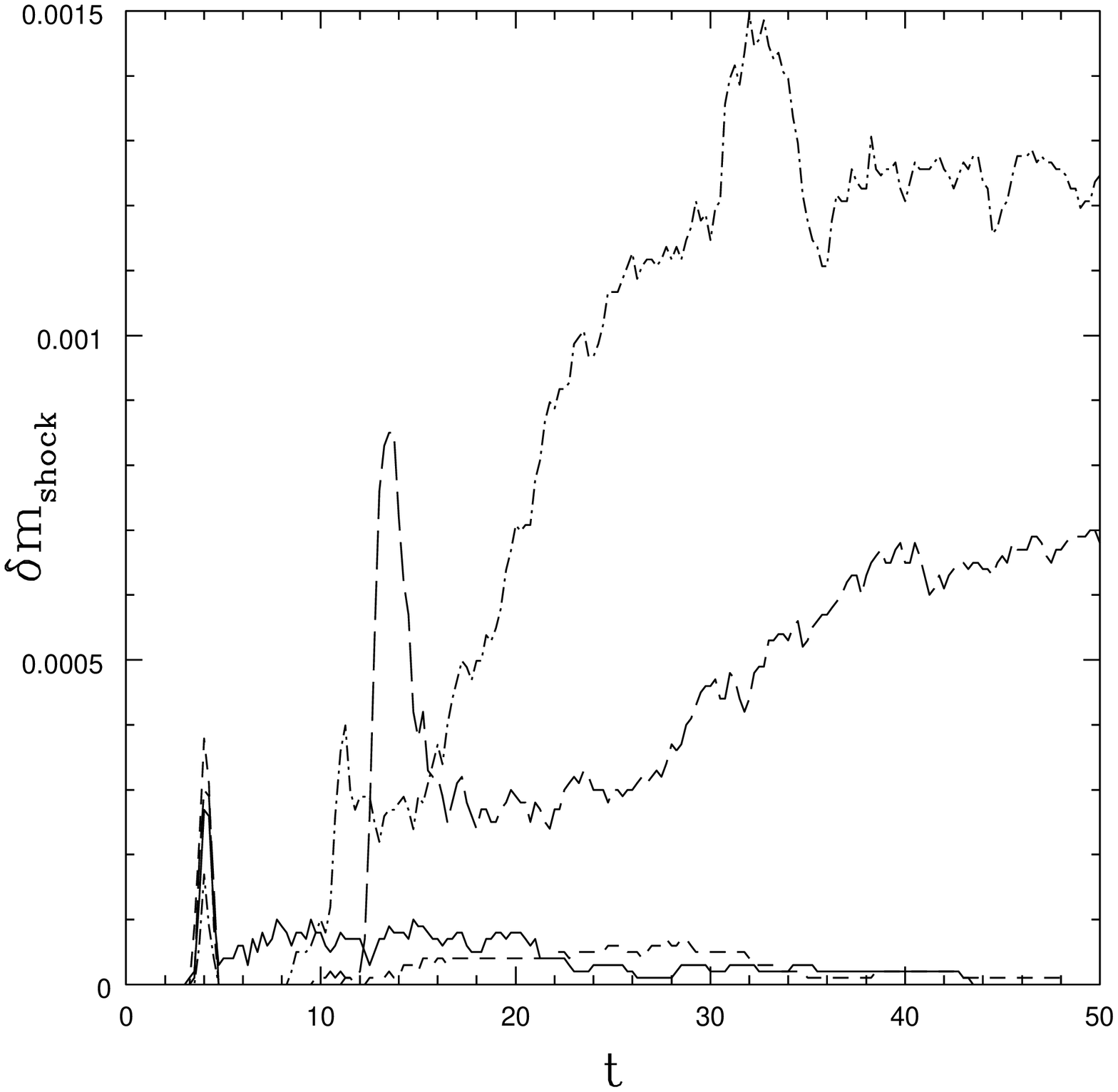,width=0.5\textwidth}}
\caption{Amount of material involved in shocks as a function of time for the
  four simulations with $\gamma=1.4$ (solid line), $\gamma=1.5$ (short-dashed
  line), $\gamma=5/3$ (long-dashed line) and $\gamma=1.8$ (dot-dashed
  line). As the polytropic index grows more mass undergoes shocks, producing
  progressively more pronouced wings in the energy distribution
  (cf. Fig. \ref{fig:specave}).}
\label{fig:deltam}
\end{figure}

Let us now look at the distribution of specific energies within the star for
the four different simulations. This is shown in Fig. \ref{fig:specave}, where
the solid lines refer to the simulations, averaged over 10 time units when the
star has reached a distance of roughly 20 times the pericentre. Note that in
each simulation, as mentioned above, the stars are somewhat inflated once they
reach pericentre. In order to compare the numerical results with the
analytical predictions we therefore have to take into account this expansion.
We have thus simply taken the initial equilibrium density as a function of
radius within the star and re-scaled the radius by a constant factor $\xi$,
thus effectively applying a homologous expansion to the stellar structure. We
have then calculated the expected energy distribution based on Eqs.
(\ref{eq:dim3}) and (\ref{eq:dim4}) for this `inflated' profile. The resulting
analytical predictions are then shown in Fig. \ref{fig:specave} with a dashed
line. The expansion factor to match the numerical data is $\xi=2.5$, 2.1, 1.63
and 1.6 for the four cases $\gamma=1.4$, 1.5, 5/3 and 1.8, respectively. It
interesting to see that this expansion parameter decreases as we increase
$\gamma$, reflecting the reduced response to variations in the gravity field
as $\gamma$ gets larger. It can be seen that for $\gamma=1.4$ our inflated
polytropic model describes very accurately the outcome of the simulation.
However, as $\gamma$ increases the results of the simulations start to deviate
from the model, in particular in the appearance of wings in the tail of the
distribution. These wings become progressively more prominent as $\gamma$ gets
larger. In the previous section we have already shown that the core of the
distribution for $\gamma=5/3$ is well described by a non inflated model.
Essentially, what is happening here is that as $\gamma$ increases the
expansion of the star becomes progressively less homologous, with more
material being pushed to higher energies (in absolute value). Since the
expansion velocity is significantly supersonic, the only way to transfer
energy within the star is through shocks, occurring in the tidal tails. As
$\gamma$ increases, the density profile of the star becomes shallower, and
more material undergoes shocks in the outer layers of the star, hence
increasing the appearance of the wings. We estimate quatitatively the
importance of shocks in our simulations in the following way. For each
particle $i$ in our simulations we compute the quantity
\begin{equation}
q_{i} = \left\{ \begin{array}{ll}
\displaystyle \frac{h_i({\mathbf{\nabla\cdot u}})_i}{c_{{\rm s},i}} &
    \mbox{when \hspace{2mm}} (\mathbf{\nabla\cdot u})_i<0 \\
                    0     &  \mbox{otherwise} 
\end{array}
\right.
\label{eq:q}
\end{equation}
where $h_i$ is the particle's smoothing length, $(\mathbf{\nabla\cdot u})_i$
is the local divergence of flow velocity and $c_{{\rm s},i}$ is the local
sound speed. The quantity $q$ is therefore non-zero and negative in regions of
convergent flow and shocks occur where $|q|\geq 1$. Fig. \ref{fig:shockimage}
shows the structure of the disrupted star for the $\gamma=5/3$ case at
$t=16.75$. The top panel shows the projected density, while the bottom panel
shows a vertical cross section of $q$. It can be seen that most of the
disrupted star is expanding, except for the tip of the tidal tails, where
there is a strong convergent flow, which has indeed $|q|>1$ and therefore
undergoes a shock.

To see how does the effect of shocks changes as the polytropic index is
varied, we also compute the quantity $\delta m_{\rm shock}$, that we define
as the total mass of particles that have $|q|>1$. Fig. \ref{fig:deltam} shows
the time evolution of $\delta m_{\rm shock}$ for the four simulations with
$\gamma = 1.4$ (solid line), $\gamma=1.5$ (short-dashed line), $\gamma=5/3$
(long-dashed line) and $\gamma=1.8$ (dot-dashed line). This plot shows a few
interesting features. First, we see that as the index $\gamma$ increases, the
amount of shocked mass increases as well, confirming our expectation that more
mass is involved with the shocks in the tidal tails. In particular, the two
simulations with the largest $\gamma$, which are the ones displaying the more
pronounced 'wings' in the energy distribution, are also the two in which more
mass undergoes shocks. Second, we see that shocks appear to occur in a
sequence of peaks. The first one, common to all simulations, occurs at
$t\approx 4$, which corresponds to pericenter passage. For the largest values
of $\gamma$, we then see a number of other peaks, which can be interpreted as
the manifestation of strongly non-linear stellar pulsations induced by the
tidal interaction (see also \citealt{ivanov01}). The period of these
oscillation decreases with increasing $\gamma$, consistent with the
expectation that the period of the fundamental mode of stellar pulsations
should vary as $(3\gamma-4)^{-1/2}$ (e.g., \citealt{cox80}).

\begin{figure*}
\centerline{\epsfig{figure=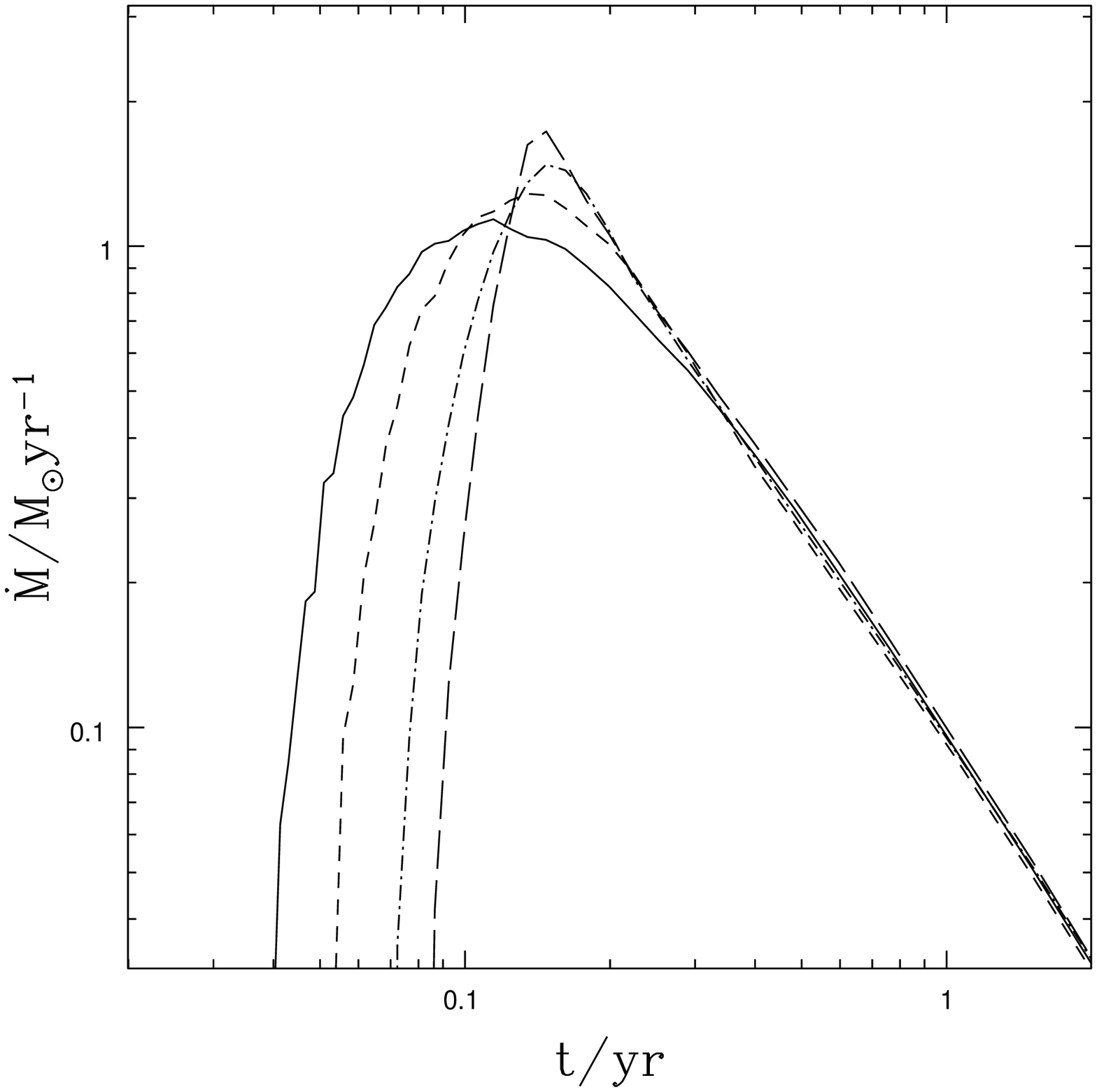,width=0.5\textwidth}
\epsfig{figure=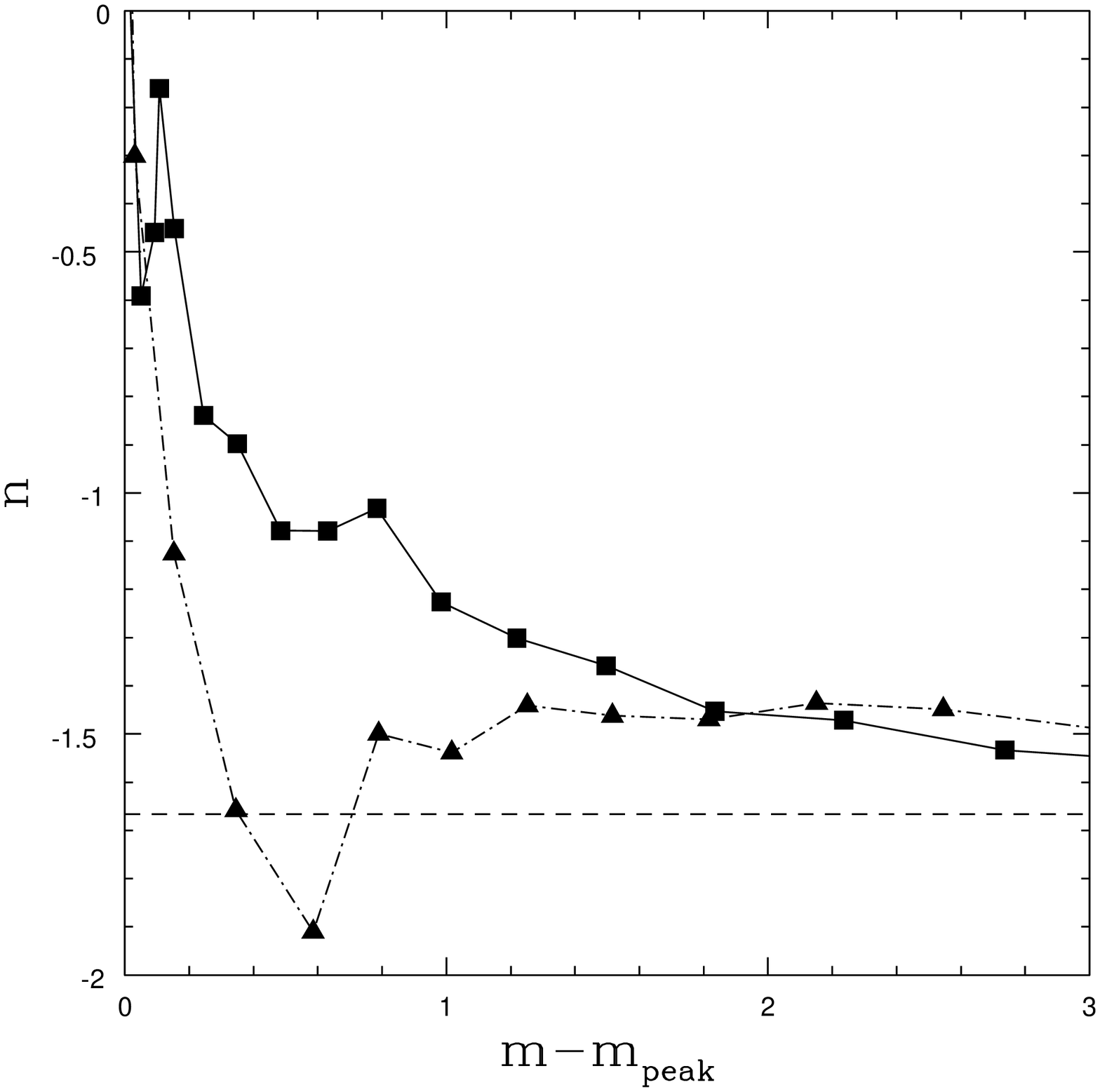,width=0.5\textwidth}}
\caption{Left: accretion rate as a function of time for the four simulations.
  Solid line: $\gamma=1.4$ (internal structure close to a solar-type star);
  short-dashed line: $\gamma=1.5$; dot-dashed line: $\gamma=5/3$ (internal
  structure appropriate for low-mass, fully convective stars); long-dashed
  line: $\gamma=1.8$. The physical units refer to the case where the disrupted
  star has $M_{\star}=M_{\odot}$ and $R_{\star}=R_{\odot}$ (note that if the
  disrupted star is a giant, the time unit is increased by a factor $(R_{\rm
    giant}/R_{\odot})^{3/2}$, which can be several hundreds, then suggesting
  that the rise to peak might be observable, and the decay time prior to
  reaching the asymptotic $t^{-5/3}$ behaviour can be very long). Right:
  instantaneous power-law index $n=\de\log\dot{M}/\de\log t$ for $\gamma=1.4$
  (squares) and $\gamma=5/3$ (triangles) as a function of magnitude drop from
  the peak. The dashed line at the bottom indicates the commonly-invoked
  power-law index $n = -5/3$ for the light curve.}
\label{fig:acc}
\end{figure*}

As mentioned in Section 2, the fact that for larger $\gamma$ the energy
distribution becomes relatively flatter implies that the distribution of
return times should become steeper and more rapidly approach the $t^{-5/3}$
profile expected for an exactly flat distribution. The resultant accretion
rate for the four simulations is shown in Fig. \ref{fig:acc} (left panel),
where the solid line refers to $\gamma=1.4$, the short-dashed line to
$\gamma=1.5$, the dot-dashed line to $\gamma=5/3$ and the long-dashed line to
$\gamma=1.8$. To give an idea of the numbers involved we have ploted the
results in physical units, assuming $M_{\star}=1M_{\odot}$ and
$R_{\star}=R_{\odot}$ (note that if the disrupted star is a giant, the time
unit is increased by a factor $(R_{\rm giant}/R_{\odot})^{3/2}$, which can be
several hundreds, then suggesting that the rise to peak might be observable,
and the decay time prior to reaching the asymptotic $t^{-5/3}$ behaviour can
be very long).  It can be indeed be easily seen that only for the largest
values of $\gamma$ does the light curve follow the $t^{-5/3}$ profile at early
time, while for lower $\gamma$ the profile gets significantly more shallow.
This is seen even better in the right panel of Fig. \ref{fig:acc}, where we
plot the instantaneous power law index $n$ (that is, the logarithmic time
derivative of the accretion rate) associated with the lightcurve for the cases
$\gamma=1.4$ (squares) and $\gamma=5/3$ (triangles) as a function of magnitude
drop from the peak (we only plot these two cases for simplicity: the two other
cases follow essentially the same behaviour). The dashed line at the bottom
indicates $n=-5/3$. This plot illustrates quite clearly that the $t^{-5/3}$
regime is only approached at late times, after the luminosity has dropped
$\sim$ 2 magnitudes from the peak for $\gamma=1.4$. In the case $\gamma=5/3$
the asymptotic regime is approached more quickly, after only a luminosity drop
of approximately 1 magnitude.

\section{Discussion and conclusions}

Candidate tidal disruption events of a star by a dormant black hole are
usually associated with luminous flares in the nucleus of an otherwise normal
galaxy. These can be detected in X-rays, for example with Chandra and ROSAT
\citep{halpern04} or with XMM \citep{esquej08}, or in the optical/UV
\citep{gezari08}. X-ray data generally observe the flare evolving down from
the peak by a few orders of magnitude, and in the best studied case, NGC 5905,
the decline appears to be consistent with a $t^{-5/3}$ fall-off
\citep{halpern04}. However, in some other cases \citep{gezari08} the
observations only span a relatively small drop in luminosity from the peak. In
these cases the lightcurve appears to be shallower than $t^{-5/3}$, and the
best fit of \citet{gezari08} to their optical data indicates a value of
$n\approx -1.1$ in one case and $n\approx -0.82$ in another. These results are
consistent with our prediction that initially the lightcurve should be
shallow, approaching a $t^{-5/3}$ profile only after the luminosity has
dropped by 2-3 magnitudes from the peak.

To summarize, in this paper we have revisited the arguments at the basis of
the expected lightcurve produced by the tidal disruption of a star in a
parabolic orbit close to a supermassive black holes. The $t^{-5/3}$ profile
originally proposed by \citet{rees88} and \citet{phinney89b} only holds in the
case where the energy distribution $\mbox{d}m/\mbox{d}\epsilon$ of the remnant
is flat, which we have shown is not the case, in general. We have proposed a
simple analytical model that relates the resultant energy distribution to the
density structure of the star. This model predicts that more centrally
concentrated (solar--type) stars should produce flares with a lightcurve
shallower than $t^{-5/3}$, approaching it only at late stages. We have tested
the model with numerical simulations and found that it does reproduce the
simulated behaviour, with the following two corrections. Firstly, we have to
account for the inflation of the star from its initial structure due to the
effective reduction of gravity as it moves in the tidal field of the black
hole. This is well described by a homologous expansion by a factor which
becomes smaller as the polytropic index becomes larger. Secondly, for large
polytropic indices we see the appearance of wings in the tails of the energy
distribution, indicating that some material has been put further away from
parabolic orbits as a result of shocks in the tidal tails.

In all cases, we do not obtain a $t^{-5/3}$ lightcurve, except at late times.
Close to the peak of the luminosity, the lightcurve is very sensitive to the
structure of the star, being shallower for stars with polytropic index close
to 4/3, expected for solar type stars. In this case, the $t^{-5/3}$ profile is
reached only after the luminosity has dropped by at least two magnitudes. For
stars with a relatively flat density profile, such as red giants and low mass
stars, the $t^{-5/3}$ profiles is reached earlier.

In this paper we have only investigated a very simple setup, with a given mass
ratio between the star and the supermassive black hole, and one given set of
orbital parameters. It is expected that the results would be further dependent
on the such additional parameters, such as the ratio of tidal radius to
pericentre distance and the eccentricity of the orbit. We plan to consider
these effects in subsequent investigations.

Finally, it should be further emphasised that all these results refer
essentially to the return time of the disrupted debris, and only correspond to
an actual luminosity under the further assumption that the subsequent
accretion is perfectly efficient and occurs on a much shorter timescale, which
may not be the case (see \citealt{ayal2000}).

\section*{Acknowledgements}

We thank Walter Dehnen for providing us with his setup routine for polytropic
spheres. We acknowledge several interesting discussions with Walter Dehnen,
Mark Wilkinson, Sergei Nayakshin and Paul O'Brien. We also thank the Referee,
Stephan Rosswog, for an insightful report. All the visualization of SPH
simulations have been obtained using the SPLASH visualization tool by Dan
Price \citep{splash}

\bibliographystyle{mn2e}

\bibliography{lodato}

\end{document}